\documentclass[aps, a4paper,superscriptaddress, nofootinbib, showpacs, twocolumn, 10pt]{revtex4}

\usepackage{amsmath,amssymb,bm,natbib}
\usepackage{epsfig}
\usepackage{graphicx}
\usepackage{slashed}

\newcommand{\beq}{\begin{equation}}
\newcommand{\eeq}{\end{equation}}
\newcommand{\bea}{\begin{eqnarray}}
\newcommand{\eea}{\end{eqnarray}}
\newcommand{\ba}{\begin{align}}
\newcommand{\ea}{\end{align}}
\newcommand{\bfig}{\begin{figure}}
\newcommand{\efig}{\end{figure}}

\newcommand{\D}{\displaystyle}

\newcommand{\gev}{\, \text{GeV}}
\newcommand{\mev}{\, \text{MeV}}

\newcommand{\tin}{t_{\rm in}}

\newcommand{\tplus}{t_{+}}

\newcommand{\omnes}{{\cal{O}}}

\begin{document}

\vskip1cm
\title{Improving the phenomenology of $K_{\ell 3}$ form factors with analyticity
and unitarity}
\author{Gauhar Abbas}
\affiliation{Centre for High Energy Physics,
Indian Institute of Science, Bangalore 560 012, India}
\author{B.Ananthanarayan}
\affiliation{Centre for High Energy Physics,
Indian Institute of Science, Bangalore 560 012, India}
\email{anant@cts.iisc.ernet.in}
\author{Irinel Caprini}
\affiliation{National Institute of Physics and Nuclear Engineering\\
POB MG 6, Bucharest, R-76900, Romania}
\author{I. Sentitemsu Imsong}
\affiliation{Centre for High Energy Physics,
Indian Institute of Science, Bangalore 560 012, India}

\begin{abstract}
The shape of the vector and scalar $K_{\ell 3}$ form factors is 
investigated by exploiting  analyticity and
unitarity in a model-independent formalism. 
The method uses as input dispersion relations for certain correlators 
computed in perturbative QCD in the deep Euclidean region,  
soft-meson theorems, and experimental information on the 
phase and modulus of the 
form factors along the elastic part of the unitarity cut. 
We derive constraints on the coefficients of the parameterizations 
valid in the semileptonic range and  on the truncation error.  
The method also predicts low-energy domains in the 
complex $t$-plane where zeros of the form factors are excluded. 
The results are useful for  $K_{\ell 3}$ data analyses
and provide theoretical underpinning for recent phenomenological
dispersive representations for the form factors.
\end{abstract}
\maketitle
\section{Introduction}
\label{sec:intro}

$K_{\ell 3}$ decays, along with the leptonic decay of the kaon,
are the gold-plated channels for a precise determination of $|V_{us}|$,
where $V_{us}$ is the  element 
of the Cabibbo-Kobayashi-Maskawa matrix (for recent reviews see \cite{Antonelli}-\cite{KaNe}). 
The amplitude of the process involves the matrix element of the 
strangeness-changing vector current between a kaon and a pion, written as: 
\begin{multline}
\langle \pi^0(p') | \overline{s}\gamma_\mu u |K^+(p) \rangle \\ =  \frac{1}{ \sqrt{2}}[(p'+p)_\mu f_+(t)+(p-p')_\mu f_-(t)],
\end{multline}
where $f_+(t)$ is the vector form factor and the combination
\beq\label{eq:f0}
f_0(t)=f_+(t)+\frac{t}{M_K^2-M_\pi^2} f_-(t)
\eeq
is known as the scalar form factor. The matrix element for the charged pion and the neutral kaon 
is related to (\ref{eq:f0}) by isospin symmetry.

The $K_{\ell3}$ decay rates were measured for the four modes 
($K = K^{\pm}$, $K^0$ and $\ell = \mu$, $e$) 
by several experimental groups \cite{Lai:2004kb}-\cite{Amsler:2008zzb}.
The rates are conveniently written as
\begin{eqnarray}
\label{eq:Mkl3}
\Gamma_{K_{\ell 3}} =\frac{ G_F^2 M_K^5}{ 192 \pi^3} C_K^2 S_{\rm EW}\, \Big|  f_+(0) V_{us} \Big|^2\,
I_K^\ell \, \left(1 + \Delta \right),
\end{eqnarray}
where $G_F$ is the Fermi constant, 
$C_{K}$ is the Clebsh-Gordan coefficient equal to $1$ ($1/\sqrt{2}$) 
for the neutral (charged) kaon decays,
$S_{\rm EW}$ is a short-distance electroweak correction, 
and $\Delta$ accounts for the electromagnetic and isospin-breaking 
corrections. The form factors enter through the value $f_{+} (0)$ at zero 
momentum transfer and  the phase space integral  $I_K^\ell$,  
which depends on the shape of the form factors in the 
physical range $M_\ell^2 \le t \le (M_K-M_\pi)^2$.

For a  precise determination of  $|V_{us}|$, it is important to improve the accuracy of  the parameterizations of the form factors 
using additional theoretical and experimental information.
Thus, the low-energy theorems based on chiral symmetry provide values of the form factors at some special points inside the 
analyticity domain \cite{AdGa}-\cite{GaLe19852}. On the unitarity cut, which extends from $t_+= (M_K+M_\pi)^2$ to $\infty$,  
the Fermi-Watson theorem \cite{Watson,Fermi:2008zz} 
implies that, below the inelastic threshold $\tin$, 
the phase is available from the corresponding partial wave of 
pion-kaon elastic scattering. 
Furthermore, recent measurements of $\tau \to K\pi \nu_\tau$ 
decays~\cite{Belle}
provide experimental information also on the modulus in the same region.

Analyticity is the ideal tool for relating the information 
from the unitarity cut to the semileptonic range. 
Several comprehensive dispersive analyses were performed recently, 
using either coupled channels  Muskhelishvili - Omn\`es 
equations \cite{JOP}-\cite{Bachirscalar2009}
or a single-channel Omn\`es representation ~\cite{Bernard:2006gy, Bernard:2009zm}. 
The dispersive representations can be extrapolated below the cut, providing
information on the shape of the form factors in the $K_{\ell 3}$ region. 
However,  direct applications of the dispersion relation
 for the data analysis are not usual , although exceptions are the Omn\`es-type relations   \cite{Bernard:2006gy,Bernard:2009zm}, 
used recently in the data analyses  by NA48 \cite{Lai2007},  KLOE \cite{Ambrosino:2007}  and KTeV \cite{Abouzaid:2009ry} 
Collaborations.  Such an analysis of BELLE data for  $\tau$ decays that probe 
the vector form factor is found in~\cite{Boito}. 

The purpose of the present paper is to discuss the implications 
of analyticity for the phenomenological analyses using an 
alternative approach proposed some years ago \cite{Okubo,SiRa,Auberson:1975ts}, known as the method of unitarity bounds. We use 
the fact that a bound 
on an integral involving the modulus squared of the form factors 
along the unitarity cut is known  from the dispersion relation 
satisfied by a certain QCD correlator.
Standard mathematical techniques then allow one to correlate 
the values of the form factor or its derivative at different points.  
For the $K\pi$ form factors, the method was applied in 
\cite{BoMaRa,Lebed,BC,Hill,AbAn,Abbas:2009dz,Abbas:2010jc}. 
The latest applications  \cite{Abbas:2009dz,Abbas:2010jc} led, in particular, to stringent constraints on  the 
shape of the scalar  form factor at low energies.

In the present work, we consider both the vector and the scalar form factors 
and focus on the phenomenological consequences of 
analyticity  and unitarity for $K_{\ell 3}$ analyses.
One of our aims is to present simple analytic constraints, easily implementable in phenomenological studies,  on the free coefficients 
of the parameterizations used in fitting the data.
In contrast to other recent works, the present work does
not require any input about the absence of zeros on the real energy line
or in the complex energy plane, nor does it require any knowledge of
the phase of the form factor in experimentally inaccessible regions.
Thus, the results of the present work are a rigorous consequence of
the general principles and do not have any model dependence.

We start by giving in Sec.~\ref{sec:unib} a brief review of 
our theoretical framework.  In Sec.~\ref{sec:input} 
we present in detail  the input quantities 
used in the application of the method to the $K_{l3}$ form factors.  
In Sec.~\ref{sec:param} we review the main parameterizations used 
in the $K_{\ell 3}$ analyses, emphasizing the merits and the 
shortcomings of each of them.  In Sec.~\ref{sec:constraints} 
we concentrate on the parameterization 
based on the standard Taylor expansion at $t=0$ and 
derive explicit constraints on the coefficients of the expansion. 
 To facilitate further applications, we present the results as simple quadratic expressions for arbitrary input values of the 
form factors at special points inside the analyticity domain (the origin $t=0$ and the Callan-Treiman(CT) point). For numerical 
illustrations, we use as input the precise values  obtained recently from calculations in ChPT and on the lattice. We work in the isospin limit, but  briefly discuss also the effects of symmetry breaking in Sec. \ref{sec:iso}. Further, in Sec.~\ref{sec:trunc} we investigate the
truncation error related to the higher order terms in the expansion, and  in
Sec.~\ref{sec:zeros} we show that the method allows one to 
derive in a rigorous way the domains in the complex $t$-plane where zeros 
of the form factors are excluded.  Sec.~\ref{Conclusion} 
contains some final remarks and our conclusions.

\section{Formalism}\label{sec:unib}
A review of the formalism of unitarity bounds-in the standard version 
and the modified forms that include additional information on the 
unitarity cut- was given recently in \cite{Abbas:2010jc}. 
Here we shall present for completeness the approach proposed in \cite{Caprini2000}, which will be used in the applications made below. 

We shall denote generically the form factors $f_+(t)$ and  $f_0(t)$  
by a function $F(t)$, which is 
real  analytic in the complex $t$-plane except for the unitarity cut 
along the positive real axis from the 
lowest unitarity branch point $\tplus=(M_K+M_\pi)^2$ to 
$\infty$. The dispersion relation satisfied by the correlator
of two strangeness-changing vector currents (see details 
in Sec.~\ref{sec:input}) implies an inequality of the type-
\beq
 \int^{\infty}_{\tplus } dt\ \rho(t) |F(t)|^{2} \leq I,
        \label{eq:I}
\eeq
where the weight function $\rho(t)\geq 0$ 
and the quantity $I$ are known.

According to the Fermi-Watson theorem \cite{Watson,Fermi:2008zz},  below the  
inelastic threshold $\tin$ the phase of $F(t)$  is equal (modulo $\pi$) to the 
phase $\delta(t)$ of a partial wave of $\pi K$ elastic scattering. Thus, we can write
\beq\label{eq:watson}
F(t+i\epsilon)= |F(t)| e^{i\delta(t)}, \quad \quad t_+<t< \tin,
\eeq
where $\delta(t)$ is known. We define the
Omn\`{e}s function
\beq	\label{eq:omnes}
 \omnes(t) = \exp \left(\D\frac {t} {\pi} \int^{\infty}_{\tplus} dt' 
\D\frac{\delta (t^\prime)} {t^\prime (t^\prime -t)}\right),
\eeq
where $\delta(t)$  is  known for 
$t\le \tin$-and is an arbitrary function, sufficiently  smooth ({\em i.e.},
Lipschitz continuous) for $t>\tin$. From  (\ref{eq:watson}) and (\ref{eq:omnes})
it follows that the function 
\beq\label{eq:h}
 h(t)= F(t) [\omnes(t)]^{-1},
\eeq
has a larger analyticity domain, namely,the complex $t$-plane cut only for $t>\tin$. 
We further assume that a reliable parameterization of the modulus $|F(t)|$ is available
  on the same range $t_+<t< \tin$, such that the quantity
\beq\label{eq:I1}
I^\prime= I - \int^{\tin}_{\tplus} {\rm d}t \rho(t) |F(t)|^2
\eeq
is known. Then from (\ref{eq:I}) we obtain an $L^2$ norm condition 
\beq\label{eq:hI}
\int_{\tin}^\infty {\rm d}t \rho(t) |\omnes(t)|^2 |h(t)|^2 \leq I^\prime
\eeq
for the function  $h(t)$  analytic in the $t$-plane cut for $t>\tin$. 
As shown in \cite{Caprini2000},  (\ref{eq:hI}) can be brought into a canonical form by making 
the conformal transformation
\beq\label{eq:ztin}
\tilde z(t) = \frac{\sqrt{\tin}-\sqrt {\tin -t} } {\sqrt {\tin}+\sqrt {\tin -t}}\,,
\eeq
which maps the complex $t$-plane cut for $t>\tin$ onto the unit disk in the $z$-plane defined by $z=\tilde z(t)$. Then,
(\ref{eq:hI}) 
can be written as
\beq\label{eq:gI1}
\frac{1}{2 \pi} \int^{2\pi}_{0} {\rm d} \theta |g(\exp(i \theta))|^2 \leq I^\prime,
\eeq
where  the function $g(z)$ is defined as
\beq\label{eq:gF}
 g(z) = w(z)\, \omega(z) \,F(\tilde t(z)) \,[O(z)]^{-1}.
\eeq 
In this relation, $\tilde t(z)$ is the inverse of $z=\tilde z(t)$, for $\tilde z(t)$ defined in (\ref{eq:ztin}) and  $w(z)$ is 
an outer function, {\it i.e.,}  a function analytic and without zeros in
$|z|<1$, whose modulus on the boundary is related to the weight $\rho(t)$
and the Jacobian of the transformation (\ref{eq:ztin}) by
\beq\label{eq:wrho}
\frac{|w(\exp(i\theta))|^2}{2\pi}=  \rho(\tilde t(\exp(i \theta) ))\, \left|\frac{{\rm d} \tilde t(\exp(i \theta))}{{\rm d}\theta}\right|,
\eeq
 In general, an outer
function is obtained from its modulus on the boundary by the integral
\beq\label{eq:wgen}
w(z)=\exp\left[\frac{1}{2\pi} \int_{0}^{2\pi} {\rm d}\theta \,
\frac{e^{i\theta}+z}{e^{i\theta}-z}\,\ln |w(e^{i\theta})| \right],
\eeq 
but in simple cases one can obtain an analytic form (see Sec. \ref{sec:outer}).
Further, the function $O(z)$ is defined as
\beq\label{eq:Otilde}
O(z) = \omnes(\tilde t(z)),
\eeq
and
\beq\label{eq:omega}
 \omega(z) =  \exp \left(\D\frac {\sqrt {\tin - \tilde t(z)}} {\pi} \int^{\infty}_{\tin} {\rm d}t^\prime \D\frac {\ln |\omnes(t^\prime)|}
 {\sqrt {t^\prime - \tin} (t^\prime -\tilde t(z))} \right).
\eeq 

From the definition (\ref{eq:gF}), taking into account (\ref{eq:h}), it follows that
 $g(z)$ is analytic within  the unit disc $|z|<1$. 
The relation (\ref{eq:gI1}) leads to constraints on the values of $g$ and its derivatives at various points 
(mathematically, this is known as the Meiman problem).
 In the general case, consider the first $K$ derivatives of $g(z)$ at $z=0$ and the values at other $N$ interior points $z_n$:
\bea\label{eq:cond}
\left[\D \frac{1}{k!} \D \frac{ d^{k}g(z)}{dz^k}\right]_{z=0}&=& g_k, \quad
0\leq k\leq K-1; \nonumber\\
 g(z_n)&=&\xi_n,  \quad z_n\ne 0, \quad  1\leq n \leq N, 
\eea
where $g_k$ and $\xi_n$ are given numbers. 
Then the following determinantal inequality holds:
\beq\label{eq:det}
\left|
	\begin{array}{c c c c c c}
	\bar{I} & \bar{\xi}_{1} & \bar{\xi}_{2} & \cdots & \bar{\xi}_{N}\\	
	\bar{\xi}_{1} & \D \frac{z^{2K}_{1}}{1-z^{2}_1} & \D
\frac{(z_1z_2)^K}{1-z_1z_2} & \cdots & \D \frac{(z_1z_N)^K}{1-z_1z_N} \\
	\bar{\xi}_{2} & \D \frac{(z_1 z_2)^{K}}{1-z_1 z_2} & 
\D \frac{(z_2)^{2K}}{1-z_2^2} &  \cdots & \D \frac{(z_2z_N)^K}{1-z_2z_N} \\
	\vdots & \vdots & \vdots & \vdots &  \vdots \\
	\bar{\xi}_N & \D \frac{(z_1 z_N)^K}{1-z_1 z_N} & 
\D \frac{(z_2 z_N)^K}{1-z_2 z_N} & \cdots & \D \frac{z_N^{2K}}{1-z_N^2} \\
	\end{array}\right| \ge 0,
\eeq
where
\beq
\bar{\xi}_n = \xi_n - \sum_{k=0}^{K-1}g_k z_n^k, \quad  \quad \bar{I} = I' - \sum_{k = 0}^{K-1} g_k^2.
\eeq
All the  principal minors of the above matrix should also be nonnegative \cite{SiRa,BoMaRa}. 

 The entries of the determinant (\ref{eq:det}) are related, by (\ref{eq:gF}),
to the derivatives $F^{(j)}(0)$, $j\le K-1$  of $F(t)$ at $t=0$, and the values
$F(t(z_n))$, respectively. It should be noted 
that (\ref{eq:det}) covers also the case of values given at complex points: if one of the numbers $t_n$ is complex, 
also the complex conjugate appears, say
$t_{n+1}=t_n^*$, 
and  we have $F(t(z_{n+1})) =F^*(t(z_{n})) $ due to the reality property. The same relations hold for 
the corresponding points in the  $z$-plane and the values $g(z_n)$.   This ensures the reality of the determinant appearing in the 
inequality (\ref{eq:det}). 

We note that the formalism presented above exploits in an optimal way the relation (\ref{eq:hI}), which is a consequence of the 
inequality (\ref{eq:I}) and of the relations (\ref{eq:watson})-(\ref{eq:I1}).
The standard approach \cite{Okubo,SiRa, BoMaRa, BC, Hill}, which does not include additional information on the form factors on 
the cut, is 
obtained formally from the above relations by setting $\tin\to\tplus$, when both the Omn\`es function $O(t)$ and the function 
$\omega(z)$ become unity. The implementation of the phase condition (\ref{eq:watson}) on the elastic cut, together 
with the relation (\ref{eq:I}), involves the solution of an integral equation of Fredholm type (for details and references, see 
\cite{Abbas:2010jc}).

 We stress that, while (\ref{eq:hI}) is a necessary condition following from the original inequality (\ref{eq:I}), it is not  
sufficient for  the fulfillment of (\ref{eq:I}), {\it i.e.}, functions that satisfy (\ref{eq:hI}) and do not satisfy (\ref{eq:I}) 
may exist. In principle, both conditions must be imposed in order to restrict the allowed domain of the parameters of interest: each condition leads to an allowed domain, the  final  region being the
intersection of the corresponding domains. As shown in \cite{Abbas:2010jc},  by a conservative choice of $\tin$, the results 
obtained from (\ref{eq:hI}) satisfy also the condition (\ref{eq:I}). We shall place ourselves in this framework here.  Finally, 
we recall that in the definition (\ref{eq:omnes}) of the Omn\`es function,  the phase for $t>\tin$ is not specified and can be 
parameterized in an arbitrary way. As shown in  \cite{Abbas:2010jc}, the results are independent of the form chosen, in particular 
of the phase at asymptotic energies, provided that the parametrization is sufficiently smooth ({\it i.e.} Lipschitz continuous).

\section{Application to the $K_{l3}$ form factors} \label{sec:input}
 In this section we  apply the above formalism to the $K_{l3}$ form factors $f_+(t)$ and $f_0(t)$. We first write down dispersion 
relations for suitable QCD correlators and show how they lead to an inequality of the type (\ref{eq:I}). Then we briefly discuss 
the low-energy theorems and the information on the phase and modulus on the elastic part of the cut used as input. For completeness 
we give also  the explicit form of the outer functions defined above.  We  work in the isospin limit, adopting the convention that 
$M_K$ and $M_\pi$ are  the masses of the charged mesons. A few comments about isospin breaking effects will be made in Sec. 
\ref{sec:iso}.

\subsection{QCD correlators}\label{sec:qcdcorrelators}
We consider the correlator of the strangeness-changing  hadronic current $V^{\nu}=\bar s\gamma^\nu u$:
\begin{eqnarray}\label{eq:ope}
\Pi^{\mu\nu}(q) &\equiv& i\int\! d^4x\, e^{iq\cdot x} \langle 0 | T\left\{ V^\mu(x) V^{\nu}(0)^\dagger \right\} | 0 \rangle  \\
&=& ( - g^{\mu\nu}q^2+q^\mu q^\nu) \Pi_1(q^2) + q^\mu q^\nu \Pi_0(q^2).\nonumber
\end{eqnarray} 
In QCD, the invariant amplitudes satisfy subtracted dispersion relations. More exactly, it is convenient to define the functions \cite{BoMaRa,BC,Hill}\footnote{The choice of the renormalization-group invariant  correlators is not unique.  Alternative definitions and the corresponding dispersion relations  are shown to lead to almost identical constraints for the $K_{\ell 3}$ form factors \cite{CaBa}.  Here we consider for convenience only the correlators (\ref{eq:chi1}) and (\ref{eq:chi0}).}
\begin{equation}\label{eq:chi1}
\chi_1(Q^2) \equiv -\frac{1}{ 2} \frac{\partial^2 }{ \partial (Q^2)^2}\left[ Q^2\Pi_1(-Q^2) \right], \end{equation}
\begin{equation}\label{eq:chi0}
\chi_0(Q^2)\equiv \frac{\partial}{ \partial Q^2} \left[ Q^2\Pi_0 (-Q^2)\right],
\end{equation}
which satisfy the dispersion relations
\begin{equation}\label{eq:chi1dr}
\chi_1(Q^2) 
=\frac {1}{ \pi} \int_0^\infty\! dt\,\frac{ t {\rm Im}\Pi_1(t) }{ (t+Q^2)^3 },  \end{equation}
\begin{equation}\label{eq:chi0dr}
\chi_0(Q^2)
= \frac{1}{\pi}\int_0^\infty\!dt\, \frac{t {\rm Im}\Pi_0(t)}{ (t+Q^2)^2}.
\end{equation}
Unitarity implies that the spectral functions are positive for $t>t_+$ and satisfy the inequalities \cite{BoMaRa,Hill}:
\beq\label{eq:unit1}
{\rm Im}\Pi_1(t) \geq \frac{3}{ 2}\frac{1}{ 48 \pi} \D \frac {\left[(t-t_+)(t-t_-)
\right]^{3/2}}{t^3} |f_+(t)|^2,
\eeq
\beq\label{eq:unit0}
{\rm Im} \Pi_0(t) \ge \frac{3}{2} \frac{t_+ t_-}{ 16\pi}
\frac{[(t-t_+)(t-t_-)]^{1/2}}{ t^3} |f_0(t)|^2,
\eeq
with $t_\pm=(M_K \pm M_\pi)^2$. 

In the limit $Q^2 >>\Lambda^2_{QCD}$, the correlators $\chi_1(Q^2)$ and $\chi_0(Q^2)$ can be calculated by perturbative QCD. 
Recent calculations to order $\alpha_s^4$  (see \cite{BCH2006, BCH2008} and references therein)   give:
\begin{multline}
\chi_1(Q^2)=\frac{1} {8 \pi^2 Q^2}
(1+\frac{\alpha_s} {\pi}-0.062 \alpha_s^2 \\ 
-0.162 \alpha_s^3-0.176\alpha_s^4),
\end{multline}
\begin{multline}
\chi_0(Q^2)=\frac{3(m_s-m_u)^2} {8 \pi^2 Q^2}
(1+1.80\alpha_s+4.65 \alpha_s^2 \\ +15.0 \alpha_s^3+57.4\alpha_s^4 ).
\end{multline}
We omitted the power corrections due to nonzero masses and QCD condensates, as they are negligible.

The relations (\ref{eq:chi1})-(\ref{eq:unit0}) show that each form factor
satisfies a relation of the type (\ref{eq:I}), where
 \bea\label{eq:rho}
\rho_+(t) = \frac{1}{ 32 \pi^2 } \D \frac {\left[(t-t_+)(t-t_-)
\right]^{3/2}}{t^2 (t+Q^2)^3 },\nonumber\\
\rho_0(t) = \frac{3\, t_+ t_- }{32 \pi^2} 
\frac{[(t-t_+)(t-t_-)]^{1/2}}{ t^2  (t+Q^2)^2},
\eea
and 
\beq\label{eq:Iqcd}
I_{+}=\chi_{1}(Q^2), \quad\quad  I_{0}=\chi_{0}(Q^2).
\eeq
We evaluated these expressions taking $Q^2 = 4 \gev^2$  as in 
\cite{BC,Hill},  $m_s(2 \gev)=98\pm 10 \mev,  m_u(2 \gev )=3\pm 1 \mev$ \cite{Lellouch} and $\alpha_s(2 \gev)=0.308\pm 0.014$, which 
results from the recent average  $\alpha_s(m_\tau)=0.330\pm 0.014$ 
\cite{Bethke}.
This gives $\chi_1(2 \gev)=(343.8 \pm  51.6 )\times 10^{-5}
\, {\rm GeV}^{-2}$ and  $\chi_0(2 \gev)= (253\pm 68)\times 10^{-6}$.

\subsection{Low-energy theorems}\label{sec:low}
The theorems based on symmetries at low energies provide useful ingredients in the applications of the above formalism.  
At $t=0$ by construction one has $f_0(0)=f_+(0)$, since $f_-(t)$ is regular at $t=0$, and SU(3) symmetry implies $f_+(0)=1$. 
Deviations from this limit
are expected to be small~\cite{AdGa} and have been 
calculated in chiral perturbation theory \cite{LeRo,BiTa} 
and more recently on the lattice   
(see the reviews in \cite{Antonelli}-\cite{KaNe}). 
In the case of the scalar form factor, 
current algebra relates the value of the scalar form factor  
at the CT point
$\Delta_{K\pi}\equiv M_K^2-M_\pi^2$ 
to the ratio  $F_K/F_\pi$ of the decay constants \cite{CallanTreiman,DaWe}:
\beq\label{eq:CT1}
f_0(\Delta_{K\pi})=F_K /F_\pi/  +\Delta_{CT}.
\eeq
To one-loop in ChPT in the isospin limit  $\Delta_{CT}= -3.1\times 10^{-3}$ \cite{GaLe19852}. Results on higher-order corrections, 
and also beyond the isospin limit, are  available \cite{KaNe, BiTa, BiGo}.

At $\bar{\Delta}_{K\pi}(=-\Delta_{K\pi})$, a
soft-kaon result \cite{Oehme} relates the value
of the scalar form factor to $F_\pi/F_K$ 
\beq\label{eq:CT2}
f_0(-\Delta_{K\pi})=F_\pi/F_K  +\bar{\Delta}_{CT}.
\eeq
A calculation  in ChPT to one-loop in the isospin limit \cite{GaLe19852} gives $\bar{\Delta}_{CT}=0.03$, but the  higher order 
ChPT corrections are expected to be  larger in this case. The estimate made in \cite{Bernard:2009zm} leads to a rather large allowed interval for $\bar{\Delta}_{CT}$.

In the present work we use as input the values of the vector and scalar form factor at $t=0$. For the scalar form factor we impose 
also the value $f_0(\Delta_{K\pi})$ at the first CT point.  As discussed in \cite{Abbas:2009dz}, due to the poor 
knowledge of $\bar{\Delta}_{CT}$, the low-energy theorem (\ref{eq:CT2}) is not useful for  further constraining the shape of 
the $K_{\ell 3}$ form factors at low energies. 

For generality, we shall present our results for arbitrary values of the parameters $f_+(0)$ and $f_0(\Delta_{K\pi})$.  For 
the numerical illustration of the results we shall use as default the values
\beq\label{eq:f0+0}
f_+(0)=0.962\pm 0.005, \quad f_0(\Delta_{K\pi})= 1.193 \pm 0.006.
\eeq
The central value of $f_+(0)$  coincides practically with the ChPT prediction given in \cite{LeRo} and is  quoted in 
\cite{Antonelli} as the most recent lattice result.  A slightly different average of lattice results $f_+(0)=0.959\pm 0.005$ is 
also quoted in \cite{Antonelli}. The value of $f_0(\Delta_{K\pi})$ is consistent with the values reported in the review 
\cite{Antonelli} in the isospin limit.

The value (\ref{eq:CT1}) at the 
first CT point is quite precise in the standard model(SM).  On the other hand, it has been suggested that 
deviations from ChPT at both CT points  would be a 
signature for physics beyond the SM, such as right-handed quark couplings to 
$W^\pm$  and charged Higgs~\cite{Bernard:2006gy,Deschamps:2009rh,Passemar}.
In what follows, we shall derive constraints on the expansion coefficients both with, and without the condition at the 
CT point. We shall also derive a relation that correlates the values of the scalar form factor at $t=0$ and 
at both CT points, which acts as an independent constraint for specific  models beyond SM.

\subsection{Phase and modulus on the  elastic region of the cut}\label{sec:phase}
We recall that the first inelastic 
threshold for the scalar form factor is set by the $K\eta$ state, and for the vector form factor by the state   $K^*\pi$, 
which suggests we take $\tin=(1\, \mbox{GeV})^2$. Strictly speaking, we must consider the inelasticity due to $K\pi\pi\pi$
at $(0.91\, \mbox{GeV})^2 $, but its influence is considered weak and may be neglected. Moreover, it is known that the elastic 
region 
extends practically up to the $K\eta'$ threshold,  which would
justify the choice $\tin=(1.4\, \mbox{GeV})^2$. 
The analysis performed in \cite{Abbas:2010jc} led to the conclusion that this choice  overconstrains the system, at least for 
the scalar form factor. Therefore, in this work we make the 
conservative choice $\tin=(1\, \mbox{GeV})^2$ for both the scalar and vector form factors.

Below $\tin$  the function $\delta(t)$ entering (\ref{eq:omnes}) 
is the phase of the $S$-wave of $I=1/2$ of the elastic $K\pi$ 
scattering for the scalar form factor, and the phase of the $P$-wave of $I=1/2$ for the vector form factor. 
In our calculations  we used as default below $\tin$ the phases  from \cite{BuDeMo,Bachirscalar2009} for the scalar form factor, 
and from \cite{Bachirvector, Bernard:2009zm} for the vector case (the differences between the two phases were taken as an estimate 
of the uncertainty related to this input; see \cite{Abbas:2009dz}).
We recall that, while the standard dispersion approaches  require a choice of the phase above $\tin$,  the present formalism is 
independent of this ambiguity \cite{Abbas:2010jc}.     Above $\tin$ we have taken  $\delta(t)$ as a smooth function 
approaching $\pi$ at high energies. As we mentioned in Sec. \ref{sec:unib}, the results are  independent of the choice of the 
phase for  $t>\tin$. 
We have checked numerically this independence with high precision.

To estimate  the low-energy integral in (\ref{eq:I1}),  we used the  Breit-Wigner parameterizations of  $|f_+(t)|$ and $|f_0(t)|$ 
in terms of the resonances given by the
Belle Collaboration  \cite{Belle} for fitting the rate of  $\tau\to K\pi\nu$ decay. This leads to the value $31.4 \times 10^{-5}\,{\rm GeV}^{-2}$ for the vector form factor and $60.9 \times 10^{-6}$  
for the scalar form factor.    By combining with the values of $I_{+,0}$ defined in  (\ref{eq:Iqcd}),  we obtain 
\beq\label{eq:Ip}
I_+'=(312 \pm 69) \times 10^{-5} \,
{\rm GeV}^{-2}, \quad I_0'=(192 \pm 90)\times 10^{-6}.
\eeq
Note that we use the Breit-Wigner parameterizations of $|f_{+,0}(t)|$  obtained in \cite{Belle}  only for estimating the low- 
energy integral appearing in (\ref{eq:I1}). The parameterizations  are not extrapolated outside the resonance region; therefore 
their 
analytic properties do not play a role in the present formalism, where the analyticity of the form factors is exactly imposed.  
We note also that the dependence on the modulus information below $\tin$  is very mild,
as it is used only for the computation of the low-energy part of
the integral (\ref{eq:I}). In fact, the results depend rather weakly on the values of $I_{+,0}'$, and moreover the 
dependence is monotonic: an increase of these quantities leads to more conservative bounds.

In our earlier work, ~\cite{Abbas:2009dz}
we had carried out an analysis on the
uncertainties to be attached to our determination associated with the
uncertainties of the inputs.  This was essential as the rather high choice
of $\tin$ of $(1.4\, \mbox{GeV})^2$. 
Having chosen for this work the conservative choice
of $(1\,\mbox{GeV})^2$, partly motivated by the considerations of our later work
\cite{Abbas:2010jc}, 
we have obtained the new constraints that comfortably accommodate
the prior results, including the uncertainties.  Therefore, we consider only the
results based on the central values of the input on the unitarity cut.

\subsection{Outer functions}\label{sec:outer}
The outer function is defined in (\ref{eq:wrho}).
Using (\ref{eq:ztin}) and (\ref{eq:rho}), a straightforward calculations leads to
\bea
 &&w_+(z)= \frac{1}{8 \sqrt{2 \pi \tin}} 
 \sqrt{1-z^2} \\&& \times \frac{(1+\tilde{z}(-Q^2))^3}
{(1-z\, \tilde{z}(-Q^2))^3} \frac{(1-z\, \tilde{z}(t_+))^{3/2} (1-z\, \tilde{z}(t_-))^{3/2}}{(1+\tilde{z}(t_+))^{3/2} 
(1+ \tilde{z}(t_-))^{3/2}}, \nonumber
\eea
for the vector form factor, and
\bea    \label{eq:ws1}
&& w_0(z)= \frac{\sqrt{3}(M_K^2-M_\pi^2)}{16 \sqrt{2\pi}\tin}
 \sqrt{1-z}\,(1+z)^{3/2}  \\
&& \times\,\frac{(1+ \tilde{z}(-Q^2))^2(1-z\,\tilde{z}(t_+))^{1/2}\,(1-z\, \tilde{z}(t_-))^{1/2}}{(1-z\,\tilde{z}(-Q^2))^2(1+ \tilde{z}(t_+))^{1/2}\,
 (1+ \tilde{z}(t_-))^{1/2}}, \nonumber
\eea
for the scalar form factor.
Here, $z$ is the current variable and $\tilde{z}(t)$ is the function defined in (\ref{eq:ztin}). The outer functions for the 
standard version \cite{BoMaRa, BC,Hill} are obtained from the above by replacing $\tin$ with $\tplus$.

\section{Parameterizations of $K_{\ell 3}$ form factors}\label{sec:param}
The first parameterizations  used simple  pole models describing 
the $t$-dependence of $f_{+}(t)$ and $f_{0}(t)$ in terms of  the lightest vector and scalar resonances with masses $M_{v}$ and $M_{s}$, respectively:
\begin{equation}
 f_{+}(t)=f_{+}(0) \,\frac{M_{v}^{2}}{M_{v}^{2}-t},\quad\quad f_{0}(t)=f_{0}(0)\,\frac{M_{s}^{2}}{M_{s}^{2}-t}.
\end{equation}
The increase of the precision of the $K_{\ell 3}$ experiments required more effective parameterizations.
\subsection{Taylor expansions }\label{sec:taylor}
The simplest expressions, adopted in practically all the experimental analyses \cite{Lai:2004kb}-\cite{Ambrosino:2007}, are  based on the Taylor expansion  around the point $t=0$:
\beq
      \hat f_+(t)=1 + \sum\limits_{k=1}^{K-1} c_{k,+} t^k,\quad  \hat f_0(t)=1 + \sum\limits_{k=1}^{K-1} c_{k,0} t^k,
       \label{eq:taylor}
\eeq
where  $\hat f_{+,0}(t)=  f_{+,0}(t)/f_{+}(0)$. The first coefficients are often expressed in terms of dimensionless slope and  curvature:
\beq\label{eq:c1c2}
c_1= \frac{\lambda'}{M_{\pi^+}^2}, \quad \quad \quad  c_2=  \frac{\lambda''}{2\, M_{\pi^+}^4}, 
\eeq
which are also related by $\lambda'= M_{\pi^+}^{2}\,\langle r^{2}_{\pi K}/6\rangle$ and  
 $\lambda''= 2 M_{\pi^+}^4\, c$  to the  radius squared $\langle r^{2}_{\pi K}\rangle$  and curvature $c$ used in some papers 
\cite{LeRo,BC,AbAn}.

The Taylor expansions converge in the disc $|t|< t_+$ limited by the first unitarity branch point.  Therefore, in the semileptonic 
range $M_l^2\leq t \leq t_-$, the convergence is expected to be rather good, with the asymptotic rate 
$t_-/t_+ = 0.31$. Of course, the convergence becomes poor  if the expansions are used outside the $K_{\ell 3}$ region.

 At the present experimental accuracy, only a few coefficients $c_k$ can be determined from the data (common choices are $K=2$ or $K=3$). A theoretical correlation between the coefficients would be helpful in fitting the data with more parameters. 
The formalism discussed here is a useful tool in this sense.   In the next section, we shall derive strong correlations between 
the slope and curvature defined in (\ref{eq:c1c2}). Also, we will show how to obtain a bound on a suitably defined truncation error, which reflects the influence of the neglected higher order terms in the expansions (\ref{eq:taylor}).

\subsection{Dispersive parameterization}\label{sec:disp}
Recently, NA48 \cite{Lai2007}, KLOE \cite{Ambrosino:2007}, 
and KTeV \cite{Abouzaid:2009ry} Collaborations reanalyzed 
their data with a dispersive representation of the Omn\`es type,   
proposed in \cite{Bernard:2006gy, Bernard:2009zm}. The  parameterizations of the two 
$K_{\ell 3}$ form factors read:
\begin{eqnarray} \label{eq:disp}
 f_{+}(t)=f_{+}(0)\exp\left[\frac{t}{M_{\pi}^{2}}(\lambda_{+}+H(t))\right],\nonumber\\
 f_{0}(t)=f_{+}(0)\exp\left[\frac{t}{M_K^{2}-M_{\pi}^{2}}(\ln[C]-G(t))\right],
\end{eqnarray}
where $\lambda_{+}$ is the slope of the vector form factor, $\ln[C]=\ln[f_{0}(M_{K}^{2}-M_{\pi}^2)]$ is the logarithm of the scalar
form factor at the CT point and the functions $H(t)$ and $G(t)$ are dispersive integrals upon the phases 
$\delta_{+,0}(t)$ of the form factors. 

The advantage of this type of parameterization is that
it includes information on the analytic 
properties of the form factors in the complex plane and on the phase, 
known (modulo $\pm \pi$) in the elastic region $t<\tin$    
from low-energy $K\pi$ phases.  However, at larger energies the phase 
is not known, and this introduces an ambiguity in the representation.
At infinity, the phase approaches a constant, whose value depends on 
the number of zeros of the form factor \cite{ACCGL, Bernard:2009zm}, so as to ensure the  asymptotic decrease like $1/t$ required by perturbative QCD \cite{Lepage}. 

Actually, the dispersion relations (\ref{eq:disp}) require only one subtraction. 
In order to  display the free parameters $\lambda_+'$  and $C$,
an additional subtraction  was performed, at $t=0$ and  $t=\Delta_{K\pi}$, respectively.  This reduces the dependence on the unknown $\delta(t)$ above $\tin$,
but at the same time spoils the asymptotic behavior of the 
form factors (see \cite{BoCaLe} for a discussion in a similar context).
The correct behavior can be restored only by imposing 
additional sum rules,
which in general are not easy to implement in the fitting procedure.

As noted above,  the representations (\ref{eq:disp}) 
assume that the form factors do not have zeros in the 
complex plane. The influence of possible zeros, 
analyzed in \cite{Bernard:2009zm}, depends  on their position 
in the complex plane.  Information about the 
presence of the zeros at low energies is therefore important for the 
dispersive representations mentioned here. In Sec. \ref{sec:zeros} 
we shall derive rigorous domains where zero values of the form factors 
are excluded.
\subsection{$z$-parameterizations}\label{sec:zparam}
A class of parameterizations used alternatively for various weak 
form factors are based on expansions in powers of a variable that 
maps conformally  the $t$-plane onto a disk. For the $K\pi$ 
form factors the method was discussed in \cite{Hill} and more recently was used by the KTeV Collaboration to reanalyze their $K_{\ell 3}$  data \cite{Abouzaid:2006}. 

The expansion is actually based  on the method of unitarity bounds discussed in the present paper.  Consider the standard version, 
based only on the inequality (\ref{eq:I}),  without information on the phase and modulus on the unitarity cut. In this case, 
as discussed at the end of  Sec. \ref{sec:unib}, one should replace  $\tin$ by $\tplus$ when $I'=I$ and the functions $O(t)$ 
and $\omega(z)$ defined in (\ref{eq:omnes}) and (\ref{eq:omega}), respectively, are equal to unity. Then, referring for 
illustration to the vector form factor,  from (\ref{eq:gF}) one can write the representation 
\beq \label{eq:zparam}
f_+(t) = \frac{1}{w_+(z)}\,\sum_{k=0}^\infty g_k z^k, 
\eeq
where 
\beq\label{eq:wplus}
w_+(z)= \frac{\sqrt{1-z^2} }{32\sqrt{\pi t_+}} 
 \frac{(1+\bar{z}(-Q^2))^3 (1-z\, \bar{z}(t_-))^{3/2}}
{(1-z\, \bar{z}(-Q^2))^3  (1+ \bar{z}(t_-))^{3/2}}.
\eeq
Here $z=\bar{z}(t)$, where
\beq\label{eq:z}
\bar{z}(t)=\frac{\sqrt{t_+}-\sqrt {t_+ -t} } {\sqrt {t_+}+\sqrt {t_+ -t}}\,.
\eeq 
An advantage of the $z$-expansion is that it allows one to derive a bound on the truncation error, describing the effect of the 
neglected higher order terms in the expansion \cite{Hill}. 
On the other hand, from (\ref{eq:wplus}) it follows that the outer function vanishes at $z=\pm 1$, points that by (\ref{eq:z}) 
correspond to $t_+$ and infinity in the $t$-plane, respectively. The zeros in the denominator are not compensated automatically 
if the  sum in the numerator of (\ref{eq:zparam}) is truncated at a finite order. Therefore, the representation (\ref{eq:zparam}) 
has unphysical singularities at the threshold $t_+$ and at infinity. These deficiencies of the standard $z$-expansion were 
discussed in  the similar case of the $B\pi$ form factor in \cite{BoCaLe}, where alternative $z$-expansions free of such 
singularities were investigated. Such parameterizations are useful and  deserve further study also for the $K_{\ell 3}$ form 
factors.

\section{ Constraints on the expansion coefficients}\label{sec:constraints}
In this section we consider the most common parameterization of the $K_{\ell 3}$ form factors based on the Taylor expansions 
(\ref{eq:taylor}). This parameterization does not include in an explicit way information on the analytic 
properties of the form factors and their behavior on the unitarity cut. However, the matematical method reviewed in 
Sec. \ref{sec:unib} allows one to derive constraints on the expansion coefficients, which follow from these properties. 
One improves in this way the quality of the expansion, which includes in an implicit way additional theoretical information.  

Using as  input  the value of $I'_+$ given in (\ref{eq:Ip}) and the phase and modulus below  $\tin=(1 \, \mbox{GeV})^2$ described in Sec. \ref{sec:input}, we obtain from (\ref{eq:det}) the following constraint on the slope $\lambda_+'$ and curvature $\lambda_+''$ of the vector form factor, for an arbitrary $f_0\equiv f_+(0)$:
\begin{multline}\label{eq:normvect}
f_0^{2}[(\lambda_+'')^2  - 0.107 \lambda_+' \lambda_+'' + 2.18 \times 10^{-4}\lambda_+''+ 2.98 \times 10^{-3} (\lambda_+')^2\\
- 1.49 \times 10^{-5} \lambda_+'+4.20 \times 10^{-8}] - 4.67 \times 10^{-7}   \leq 0.
\end{multline}
The numerical coefficients of this relation depend on the phase  in the elastic region and the coefficient $I_+'$ defined in (\ref{eq:Ip}), which gives the last term in the left hand side of the inequality.

\begin{figure}[thb]
 	\begin{center}\vspace{0.5cm}
 	 \includegraphics[width = 7.3cm]{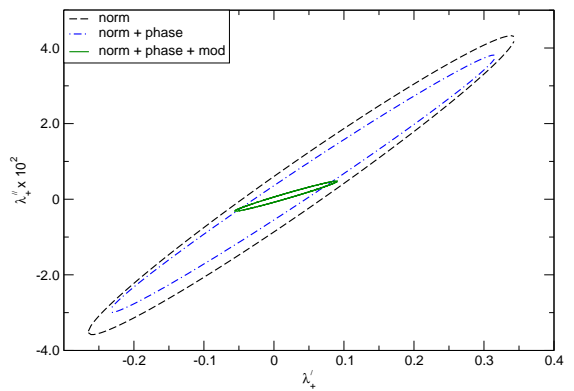}
	\caption{Allowed domain for the slope and curvature of the vector form factor, using  the normalization $f_+(0)=0.962$ and phase and modulus information up to $\tin=(1 \, \mbox{GeV})^2$.}
	\label{fig:vect1}
 	\end{center}\vspace{0.5cm}
\end{figure}

As an illustration, for the input value $f_+(0)=0.962$ given in (\ref{eq:f0+0}), the inequality (\ref{eq:normvect}) is represented 
as  the interior of the smallest ellipse in the slope-curvature  plane in Fig. \ref{fig:vect1}. The alternative input  
$f_+(0)=0.959$ adopted in \cite{Antonelli} leads to practically the same ellipse. For completeness, we represent in the same 
figure the domains obtained using as input the normalization at $t=0$ and the standard unitarity bounds, 
without information on the phase and modulus 
(the largest ellipse), 
and the domain obtained by including in the standard unitarity bounds the phase on the elastic region, known from the Fermi-Watson theorem 
(the intermediate ellipse). The small ellipse is situated inside the other two, 
which confirms that all the constraints are satisfied by the domain 
described by the inequality (\ref{eq:normvect}). 

\begin{figure}[t]
 	\begin{center}\vspace{0.1cm}
 	 \includegraphics[width = 7.3cm]{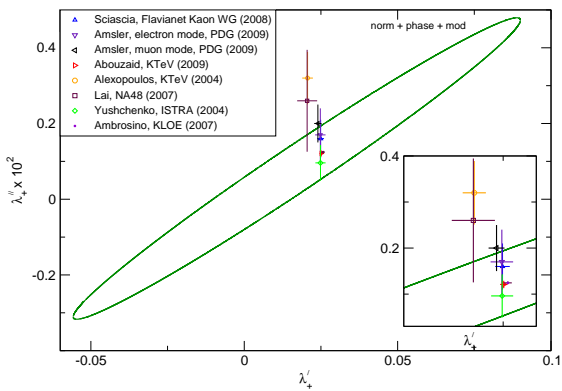}
	\caption{The best constraints for the slope and curvature of the vector form factor for $f_+(0)=0.962$ compared with experimental determinations.}
	\label{fig:vect2}
 	\end{center}\vspace{0.5cm}
\end{figure}

In Fig. \ref{fig:vect2}, the constraint (\ref{eq:normvect}) is represented together with experimental points from 
\cite{Yushchenko:2003xz,arXiv:0812.1112,Amsler:2008zzb,Ambrosino:2007,Abouzaid:2009ry,Alexopoulos:2004sy,Lai2007}, where  we have extracted the 
corresponding curvature from the constrained
fit given in ~\cite{Ambrosino:2007}. 
We note that, except the results from NA48 and KLOE, which have curvatures slightly larger than the allowed values, 
the experimental data satisfy the constraints.  We note also that the theoretical predictions $ \lambda_+'= (24.9 \pm 1.3)\times 
10^{-3}$, $\lambda_+''= (1.6 \pm 0.5)\times 10^{-3} $  obtained from ChPT to two loops \cite{KaNe}, and 
$ \lambda_+'= (26.05_{-0.51}^{+0.21})\times 10^{-3}$, $\lambda_+''= (1.29_{-0.04}^{+0.01})\times 10^{-3} $   \cite{Bachirvector}, 
and $ \lambda_+'= (25.49\pm0.31)\times 10^{-3}$, 
{\bf $\lambda_+''= (1.22\pm0.14)\times 10^{-3} $ \cite{Boito} } obtained from 
dispersion relations are consistent  with the constraint: the expression in the left side 
of (\ref{eq:normvect}) is negative when evaluated with $f_+(0)=0.962$ and the central values of the slope and curvature 
given above.

\begin{figure}[thb]
 	\begin{center}\vspace{0.5cm}
 	 \includegraphics[width = 7.3cm]{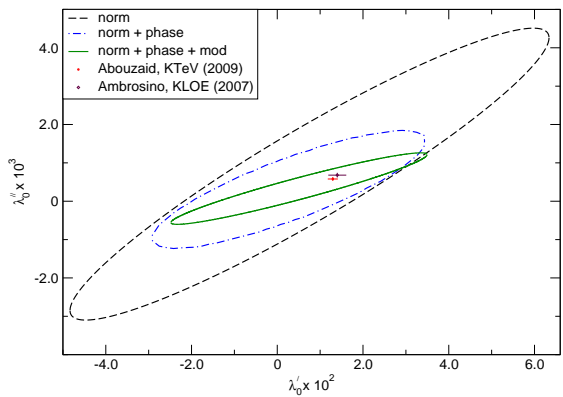}
	\caption{Allowed domain for the slope and curvature of the scalar form factor, using as input the normalization $f_+(0)=0.962$ and phase and modulus information up to  $\tin=(1 \, \mbox{GeV})^2$. }
	\label{fig:scal1}
 	\end{center}\vspace{0.5cm}
\end{figure}

The analogous constraint for the slope and curvature of the  scalar form factor  for an arbitrary normalization $f_0\equiv f_+(0)$ reads:
\begin{multline}\label{eq:normsc}
f_0^{2}[(\lambda_0'')^2  - 0.059 \lambda_0' \lambda_0'' - 3.58 \times 10^{-4}\lambda_0'' + 9.72 \times 10^{-4} (\lambda_0')^2\\  
+ 9.64 \times10^{-6} \lambda_0'+3.67 \times 10^{-8}] -8.05 \times 10^{-8}   \leq 0.
\end{multline}
In deriving this constraint, we used the value of $I'_0$ from (\ref{eq:Ip}) and the phase and modulus described in Sec. 
\ref{sec:input}.  For illustration, for $f_0=0.962$  we obtain the small ellipse in Fig. \ref{fig:scal1} where, as in 
Fig. \ref{fig:vect1},  the larger ellipses are obtained using the standard unitarity bounds. These constraints are  satisfied by 
the points in the domain (\ref{eq:normsc}).

If we include in addition the constraint at the first CT point, $f_{CT}\equiv f_0(\Delta_{K\pi})$,  we obtain the inequality
\begin{multline}\label{eq:normscCT}
f_0^{2}\,[(\lambda_0'')^2  + 0.25 \lambda_0' \lambda_0'' + 21.6 \times 10^{-3}\lambda_0'' + 15.3 \times 10^{-3} (\lambda_0')^2 \\
+2.68 \times 10^{-3} \lambda_0'+1.17 \times 10^{-4}] - 10^{-3} f_0 f_{CT} 
( 2.67 \lambda_0' \\+ 21.53 \lambda_0''+0.23)
+1.16 \times 10^{-4} f_{CT}^2 
-3.23 \times 10^{-10}   \leq 0.
\end{multline}
For the central values given in (\ref{eq:f0+0}), this domain is the interior  of the small ellipse in Fig. \ref{fig:scal2}. 
The large ellipse in this figure represents the allowed domain obtained from the standard unitarity bounds with the 
same input at interior points, and the intermediate ellipse is obtained from the standard unitarity bounds by 
imposing also the phase below $\tin$ according to the Fermi-Watson theorem. The small ellipse is situated inside 
the other two, which shows that the slope and the curvature which satisfy the inequality (\ref{eq:normscCT}) obey 
also the standard unitarity bounds.

The above domains were obtained for the central values of the parameters in (\ref{eq:f0+0}). 
As shown  in \cite{Abbas:2009dz}, the inclusion of uncertainties in the inputs has the effect of slightly 
 enlarging the allowed domains. In particular, there is a straightforward dependence
of the shape of the ellipses on the input $I'$.  These could be subject to uncertanties both in pQCD as well 
as due to the uncertainties in  modelling the modulus information.  However, the resulting ellipses
simply shrink or expand if $I'$ is taken to be smaller or larger.

We note that the theoretical prediction of ChPT to two loops 
$ \lambda_0'= (13.9_{+1.3}^{-0.4}\pm0.4)\times 10^{-3}$, $\lambda_0''= (8.0_{+0.3}^{-1.7})\times 10^{-4} $ 
reported in \cite{KaNe} is consistent within errors with the constraint (\ref{eq:normscCT}) with the default 
input (\ref{eq:f0+0}): for the central value of the slope $ \lambda_0'$ given above, the range of  
$\lambda_0''$ allowed by (\ref{eq:normscCT}) is $(8.24 \times
10^{-4}, 8.42\times 10^{-4})$. The same is true for the theoretical 
prediction $ \lambda_0'= (16.00 \pm 1.00 )\times 10^{-3}$, $\lambda_0''= (6.34\pm 0.38)\times 10^{-4} $ 
obtained in \cite{Jamin:2004re} from dispersion relations.

As concerns the experimental values, Figs.  \ref{fig:scal1} and \ref{fig:scal2} show that the determinations 
\cite{Ambrosino:2007}
(we have extracted the curvature from the
constrained fit therein) and \cite{Abouzaid:2009ry} are consistent with the phase and modulus information together with the normalization (\ref{eq:f0+0}),  but are 
outside the domain obtained when we impose also the CT theorem, with the input value given in (\ref{eq:f0+0}). 

\begin{figure}[htb]
 	\begin{center}\vspace{0.4cm}
 	 \includegraphics[width =7.3cm]{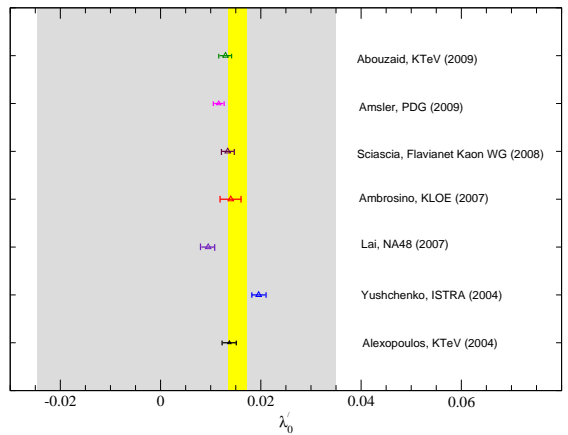}
	\caption{Allowed bands for the slope of the scalar form factor obtained from (\ref{eq:normsc}) large band, and (\ref{eq:normscCT}) 
 narrow band, for $f_+(0)=0.962$ and $f_0(\Delta_{K\pi})=1.193$, compared with the experimental information.}
	\label{fig:scal3}
 	\end{center}\vspace{0.5cm}
\end{figure}

In Fig. \ref{fig:scal3}, we  compare the allowed bands for the slope $\lambda_0'$,  corresponding to the end points 
of the ellipses defined by the inequalities (\ref{eq:normsc}) and (\ref{eq:normscCT}) for the same input as above, 
with the experimental determinations.  As noted already, the slope predicted by NA48 \cite{Lai:2007dx} is not 
consistent with the SM input at $t=0$ and $t=\Delta_{K\pi}$. This conclusion is very stable with respect 
to the input value of $I_0'$, which gives the last term in the l.h.s. of (\ref{eq:normscCT}): it turns out 
that only by increasing this value by a factor of almost 5, the allowed ellipse is inflated enough as to 
include the central value of the slope from \cite{Lai:2007dx}.  The relation (\ref{eq:normscCT}) implies 
also that, keeping $I_0'$ and $f_+(0)$ fixed, one must reduce the  input value at the CT point down to  $f_0(\Delta_{K\pi})=1.138$ in order to have the central value of the slope from \cite{Lai:2007dx} inside the allowed domain. 

We end this section on the shape of the $K_{\ell 3}$ form factors with a  comment on the low-energy theorem 
(\ref{eq:CT2}) at the second CT point. As discussed in Sec. \ref{sec:low},  due to the poor knowledge of 
the correction $\bar{\Delta}_{CT}$, this  theorem does not constrain  further  the coefficients of the expansion
 beyond the domain obtained with the input at $t=0$ and $t=\Delta_{K\pi}$. On the other hand, using these values as input , the formalism presented in Sec. \ref{sec:unib} leads to an allowed domain for the value of the scalar form factor at  $t=-\Delta_{K\pi}$. Namely, from the proper input in the determinant  (\ref{eq:det}) we obtain the inequality
\begin{multline}\label{eq:ctctbar}
 0.177 \bar{f}_{CT}^2 + 0.086 f_{CT}^2 +  0.523 f_0^2 - 
    0.425 f_0 f_{CT}  - \\
    0.608 f_0  \bar{f}_{CT}+ 0.246 f_{CT} \bar{f}_{CT}
   -1.92 \times 10^{-4} \leq 0,
\end{multline}
where $\bar{f}_{CT}\equiv f_0(-\Delta_{K\pi})$. This relation correlates the values of the 
scalar form factor at $t=0$ and at the two CT points and represents a nontrivial anayticity constraint for the predictions beyond the SM.  As shown in \cite{Abbas:2009dz}, for the  input values  (\ref{eq:f0+0}), the inequality (\ref{eq:ctctbar}) leads to a narrow range $-0.046 \leq \bar{\Delta}_{CT} \leq 0.014$ for the correction defined in (\ref{eq:CT2}),  improving the ChPT prediction reported in \cite{Bernard:2009zm}.

\subsection{Isospin breaking effects}\label{sec:iso}
 The bounds given above were obtained in the limit of isospin symmetry. We recall that in our convention $M_K$ and $M_\pi$ are 
the masses of the charged mesons. As discussed recently \cite{KaNe}, the isospin breaking effects should be taken 
into account at the present level of experimental and theoretical precision. At low energy, at next-to-leading order 
 in the chiral expansion, there are strong isospin violations at order $(m_d-m_u)p^2$ 
and electromagnetic corrections at order $e^2 p^2$, while the 
next-to-next-to-leading order terms include corrections up to 
order $(m_d-m_u)^2 p^4$.  

In the formalism  considered here, isospin symmetry was used in the unitarity relations (\ref{eq:unit1}) and (\ref{eq:unit0}), where  the matrix elements of the $ K^+\pi^0$ and $ K^0\pi^+$ pairs entering the unitarity sum were related by symmetry. The description of the form factors on the unitarity cut in terms of resonances suggests that the isospin corrections in this region are small (for a detailed discussion see \cite{Bachirvector}).  So, the unitarity relations (\ref{eq:unit1}) and (\ref{eq:unit0})  and, consequently, the outer functions defined in Sec. \ref{sec:outer} conserve their form. We checked also that a change of about 0.02 rad of the phase below $\tin$, suggested to represent isospin effects \cite{Bernard:2009zm}, leaves the results practically unmodified. 

If the symmetry is broken, one must use  the physical pion and kaon masses in the evaluation of the unitarity threshold and the phase space factors. To illustrate this effect, we recalculated the constraints (\ref{eq:normvect})-(\ref{eq:normscCT}) using  the same input on the unitarity cut, with the exception of $M_\pi$, taken to be the mass of the neutral pion. 
For instance,  we obtain instead of (\ref{eq:normscCT}) the inequality
\begin{multline}\label{eq:normscCTiso}
 f_0^{2}\,[(\lambda_0'')^2  + 0.24 \lambda_0' \lambda_0'' + 21.3 \times 10^{-3}\lambda_0'' + 15.1 \times 10^{-3} (\lambda_0')^2 \\
+2.62 \times 10^{-3} \lambda_0'+1.14 \times 10^{-4}] - 10^{-3} f_0 f_{CT} 
( 2.61 \lambda_0' \\+ 21.23 \lambda_0''+0.23)
+1.13 \times 10^{-4} f_{CT}^2 
- 3.21 \times 10^{-10}   \leq 0,
\end{multline}
where the  definition of the slope and curvature is still based on the mass of the charged pion, as in (\ref{eq:taylor}).  
The small differences between  (\ref{eq:normscCT}) and (\ref{eq:normscCTiso}) are due to the fact that the position of 
the CT point and its image in the $z$-plane are slightly changed if the pion mass is modified.   

\begin{figure}[htb]
 	\begin{center}\vspace{0.1cm}
 	 \includegraphics[width = 7.3cm]{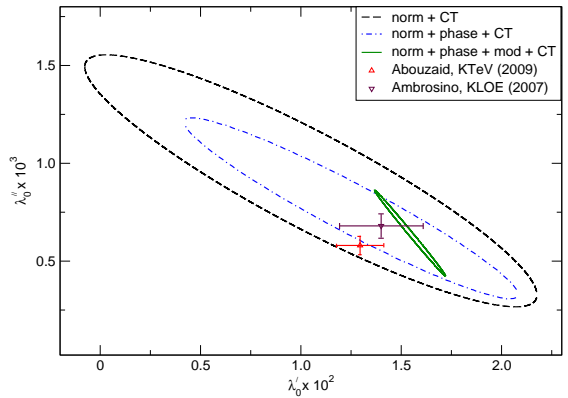}
	\caption{  Allowed domain for the slope and curvature of the scalar form factor, using the normalization $f_+(0)=0.962$, the value $f_0(\Delta_{K\pi})=1.193$,  and phase and modulus information up to  $\tin=(1 \, \mbox{GeV})^2$.}
	\label{fig:scal2}
 	\end{center}\vspace{0.5cm}
\end{figure}

The inequality  (\ref{eq:normscCTiso}) can be used to constrain the  slope and curvature for the $K^+\pi^0$ form factor 
using the isospin corrections for the same form factor at $t=0$ and the CT point \cite{BiGo,KaNe}. For instance, by increasing simultaneously the values given in (\ref{eq:f0+0}),  $f_+(0)$  by 2\% \cite{BiGo} and $f_0(\Delta_{K\pi})$ by 0.029 \cite{BiGo,KaNe}, we obtain from  (\ref{eq:normscCTiso}) an ellipse slightly shifted towards the right by the amount $\delta \lambda_0' \approx 0.0007$ compared to the small ellipse in Fig. \ref{fig:scal2}. The shifted ellipse leads to curvatures slightly higher for the same slope than those obtained in the isospin limit.  

\vskip0.4cm

\section{Bounds on the truncation error}\label{sec:trunc}

One might ask  how accurate is the representation of the $K_{\ell 3}$ form factors in the physical region by a few 
terms in the Taylor expansion (\ref{eq:taylor}). It is also of interest to know the extrapolation error if 
the truncated  expansion is used beyond the physical region up to,  say, the CT point $t=\Delta_{K\pi}$. To answer 
these questions one needs an estimate of the  higher-order terms neglected in usual parameterizations. 
As discussed in previous works, for instance \cite{Lebed}, \cite{Hill}, \cite{BoCaLe}, the technique 
of unitarity bounds allows one to find a model-independent estimate  of the theoretical error produced
 by the truncation of the Taylor expansion.

We assume that  the fit provides values of the coefficients 
$c_k$, for $k \leq K-1$, in the truncated  expansion (\ref{eq:taylor}). Since the series is convergent in the physical region, one expects the 
influence of the higher terms to be gradually smaller. We recall that, asymptotically, the convergence rate scales as $t/t_+$, where $t_+$ is the convergence radius, and this ratio does not exceed 0.31 in the semileptonic region.

Following, for instance, the suggestion made in \cite{BoCaLe}, one can  define   the truncation error as the first term neglected in the expansion: 
\beq\label{eq:deltaf}
\delta \hat f_{+,0}(t)_{\rm trunc} \sim |c_K t^K|.
\eeq
As  further suggested  in \cite{BoCaLe}, it is reasonable to increase the number of terms until the truncation error becomes smaller that the experimental one.

Without additional information, it is in general impossible to estimate the magnitude of higher terms in an expansion, even if it 
is convergent. Fortunately, in the present case, using the fundamental inequality (\ref{eq:det}) of Sec. \ref{sec:unib} and the 
relation (\ref{eq:gF}) between the
function $g$ and the form factor $F$, it is possible to derive upper and lower bounds on the next coefficient  $c_K$  in terms of the 
coefficients  $c_k$,  supposed to be known for $k \leq K-1$.

For illustration, we give below the constraint relating the value $f_0\equiv f_+(0)$, the slope $\lambda'$ 
and the curvature $\lambda''$, to the next coefficient $c_3$. For the vector form factor the relation is
\begin{multline}\label{eq:c3vect}
f_0^2[c_3^2 +c_3 (-0.045 +79.2 \lambda_+' -4953.4 \lambda_+'')\\ +1892.5 (\lambda_+')^2 - 3.39\lambda_+'- 2.07 \times 10^{5}  \lambda_+' \lambda_+''
+134.45  \lambda_+''\\ +6.24 \times 10^{6} (\lambda_+'')^2 +5.1 \times 10^{-3}] -0.051 \leq 0,
\end{multline}
while the similar relation for the scalar form factor, without imposing the CT theorem,  reads
\begin{multline}\label{eq:c3sc}
f_0^2[c_3^2 +c_3 (-0.29 +8.41 \lambda_0' - 3321.5 \lambda_0'')\\ 
+ 123.2 (\lambda_0')^2 - 0.18 \lambda_0'-2.04 \times 10^{4} \lambda_0' \lambda_0''
+445.9 \lambda_0''\\ + 2.87 \times 10^{6} (\lambda_0'')^2 +0.025] -0.0087\leq 0.
\end{multline}
The numerical coefficients in these inequalities depend on the particular dispersion relation satisfied by the QCD correlators, their perturbative expressions, and the information on the phase and modulus on the elastic part of the cut.

As a numerical example, let us take for the vector form factor the central values  $\lambda_+'=25.09 \times 10^{-3}$ 
and $\lambda_+''=1.21 \times 10^{-3}$ \cite{Abouzaid:2009ry}. Then, for our choice   $f_0(0)=0.962$, we obtain from (\ref{eq:c3vect}) the allowed interval $1.79 \,\mbox{GeV}^{-6} 
\le c_3 \le 2.25 \,\mbox{GeV}^{-6}$, which implies a correction at the end of the semileptonic region 
$\delta \hat f_+(t_-)$ between $3.5 \times 10^{-3}$ and $4.4 \times 10^{-3}$. 

For the scalar  form factor, taking a point $\lambda_0'= 15 \times 10^{-3}$ and $\lambda_0''= 0.69\times 10^{-3}$, situated inside the 
small ellipses in Figs. \ref{fig:scal1} and \ref{fig:scal2},  we obtain from (\ref{eq:c3sc}) the range  $1.14 \,\mbox{GeV}^{-6} \le c_3 
\le 1.32 \,\mbox{GeV}^{-6}$, which implies a correction  $\delta \hat f_0(t_-)$ at the end of the semileptonic region
between $2.2 \times 10^{-3}$ and $ 2.5 \times 10^{-3}$. 
On the other hand, evaluated  at the CT point, the term $c_3 t^3$ produces a correction  of about 
0.014 to  $f_0(\Delta_{K\pi})$,  larger than the error quoted in (\ref{eq:f0+0}). So, even for a quadratic 
parameterization of the scalar form factor, the extrapolation to the CT point is not precise enough for testing possible deviations from the SM. Of course,  in practical applications one should use in (\ref{eq:c3vect}) and  (\ref{eq:c3sc}) the values of the slope and  curvature obtained from the fits of the data. 

We note  that the relations (\ref{eq:c3vect}) and  (\ref{eq:c3sc})  are useful also if one uses a  cubic parameterization 
in the experimental analysis. In this case, the relations provide theoretical constraints for the next coefficient $c_3$ 
included in the fit. The truncation error can then be estimated using a constraint on the next coefficient $c_4$, which 
is obtained in a straightforward way from the general relation (\ref{eq:det}).

\section{Domains where zeros are excluded}\label{sec:zeros}
The question of the zeros of the form factors  is important from theoretical and practical points of view. The dispersive 
representations of the Omn\`es type require the knowledge  of the zeros in the complex plane.  The zeros are 
important also in  ChPT, where  their  presence is  required in some cases by symmetry  arguments \cite{Leutwyler:2002hm, ACCGL}. 

A study of the zeros of the pion electromagnetic form factor was performed in \cite{Raszillier:1976as}. For the $K\pi$ form 
factors, the influence of possible zeros in the context of Omn\`es dispersive representations has been
analyzed in \cite{Bernard:2009zm}. The absence of zeros is assumed in the recent analysis of KTeV data reported in \cite{Abouzaid:2009ry}. 

 The mathematical techniques presented in Sec. \ref{sec:unib} can be adapted in a straightforward way  to  the problem of zeros.  Let us assume that 
the form factor $F(t)$ has a simple zero on the real axis, $F(t_0)=0$. From the relation (\ref{eq:gF}) it follows that 
$g(z_0)=0$, where $z_0=\tilde z(t_0)$. We shall use this information in the determinant condition (\ref{eq:det}): if the zero 
is compatible with the remaining information, the inequality (\ref{eq:det}) can be satisfied. If, on the contrary, the inequality 
is violated, the zero is excluded. It follows that we can obtain from (\ref{eq:det}) a rigorous condition for the domain of points 
$z_0$ (or $t_0$) where the zeros are excluded. First assume that we use as input only the value of the form factor at $t=0$. 
Then from (\ref{eq:det}) the domain is given by
\beq	\label{eq:detz1}
\left|
	\begin{array}{c c }
	I'-g_{0}^{2} & -g_0\\	
	-g_0 & \frac{\tilde z(t_0)^2}{1-\tilde z(t_0)^2}  \\
	\end{array}\right|\leq 0.
\eeq
Here, $I'$ is defined in (\ref{eq:I1}), $g_0$ is related to the value $F(0)$ by the relation (\ref{eq:gF}), and $\tilde z(t)$ is defined in (\ref{eq:ztin}). If we include in addition the value of the form factor
at some point $t_1$ (for instance,  $t_1=\Delta_{K\pi}$ for the scalar form factor), 
the condition reads:

\beq	\label{eq:detz2}
\left|
	\begin{array}{c c c }
	I'-g_{0}^{2} & g(\tilde z(t_1))-g_0 & -g_0\\	
	g(\tilde z(t_1))-g_0 &  \frac{\tilde z(t_1)^2}{1-\tilde z(t_1)^2} & \frac{\tilde z(t_1)\tilde z(t_0)}
  {1-\tilde z(t_1)\tilde z(t_0)}\\
	-g_0 &\frac{\tilde z(t_1)\tilde z(t_0)}{1-\tilde z(t_1)z(t_0)} & \frac{\tilde z(t_0)^2}{1-\tilde z(t_0)^2}  \\
	\end{array}\right|\leq 0.
\eeq
To illustrate the method we use the default input $f_+(0)=0.962$ and $f_0(\Delta_{K\pi})=1.193$.   Then, from (\ref{eq:detz1}) 
it follows that in the case of the vector form factors, simple zeros are excluded in the interval $-0.31\, \mbox{GeV}^2 \leq 
t_0 \leq 0.23\, \mbox{GeV}^2$ of the real axis, while for the scalar form factor  the range with no zeros is  
$-0.91 \, \mbox{GeV}^2\leq t_0 
\leq 0.48 \mbox{GeV}^2$. 
If we also impose the low-energy theorem (\ref{eq:CT1}), with the value of $f_0(\Delta_{K\pi})$ from (\ref{eq:f0+0}),
the condition (\ref{eq:detz2}) implies that the scalar 
form factor  cannot have simple zeros  in the range  
$-1.81  \, \mbox{GeV}^2 \leq t_0 \leq 0.93 \, \mbox{GeV}^2$.
The formalism  rules out zeros in the
physical region of the kaon semileptonic decay.

We have also studied the sensitivity of the variation of our inputs.
The dependence on the parametrizations of the scattering phase shifts;
the uncertainties in $f_+(0)$ and $f_0(\Delta_{K\pi})$ are found to be 
imperceptible.
On the other hand, the uncertainties on the quantity $I'_+$ given in (\ref{eq:Ip}) lead to
the following limits for the region where zeros are excluded for
the vector form factors: $-0.28 \, \mbox{GeV}^2 \leq t \leq 0.22 \mbox{GeV}^2$ for the maximum value
and $-0.36\, \mbox{GeV}^2 \leq t \leq 0.26 \mbox{GeV}^2$ for the minimum value.  
The corresponding limits for the scalar form factor upon
inclusion of the constraint from the CT points
from the uncertainties in $I'_0$, quoted in  (\ref{eq:Ip}), are: $-1.60\, \mbox{GeV}^2 \leq t \leq 0.91 \mbox{GeV}^2$ and
$-2.26\, \mbox{GeV}^2 \leq t \leq 0.97 \mbox{GeV}^2$ respectively.

\begin{figure}[htb]
 	 \includegraphics[width = 7.1cm]{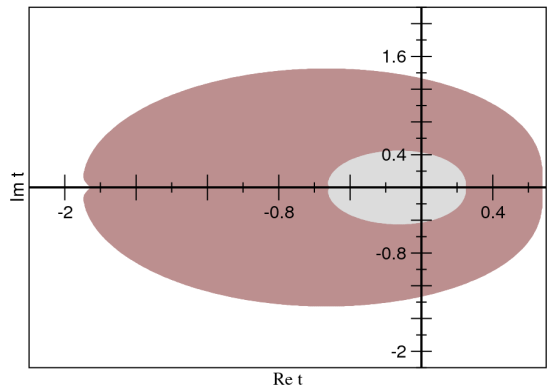}
	\caption{Domain where zeros of the vector form factor are excluded,  derived from (\ref{eq:detz3}). The small domain is 
obtained without including phase and modulus in the elastic region.}
	\label{fig:dom1}
\end{figure}

\begin{figure}[htb]
 	 \includegraphics[width = 7.1cm]{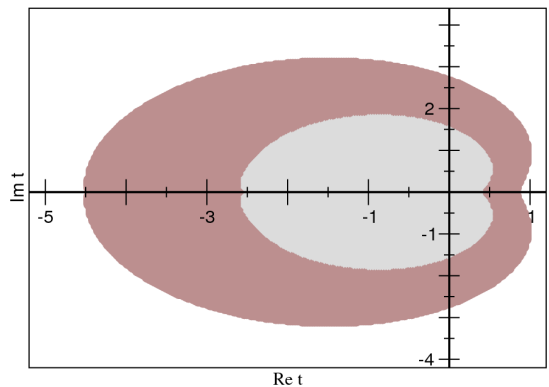}
	\caption{Domain without zeros for the scalar form factor, derived from (\ref{eq:detz3}) for $f_+(0)=0.962$. The small domain is obtained 
without including phase and modulus in the elastic region.}
	\label{fig:dom2}
\end{figure}

\begin{figure}[htb]
 	 \includegraphics[width = 7.1cm]{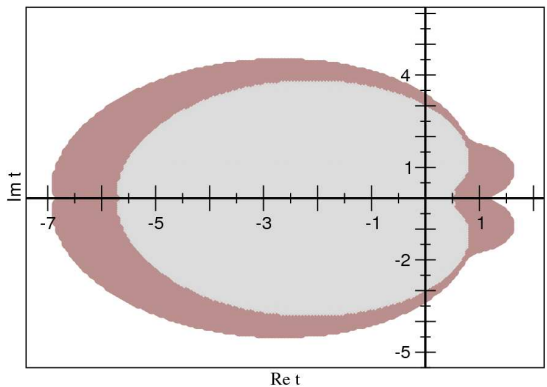}
	\caption{As in Fig. \ref{fig:dom2}, using in addition  the input $f_0(\Delta_{K\pi})=1.193$.}
	\label{fig:dom3}
\end{figure}

The method can be easily extended to the case of complex zeros. Since the functions are real analytic, {\em i.e.} 
they satisfy the 
relation 
$F(t^*)=F^*(t)$, the zeros appear in complex conjugate pairs: if $F(t_0)=0$, then  also  $F(t_0^*)=0$. We can implement this 
condition  by formally setting in (\ref{eq:detz2}) $t_1\to t_0^*$ and $g(\tilde z(t_1))\to 0$. So, the complex domain where zeros 
are excluded by the normalization at $t=0$ is given by 

\beq	\label{eq:detz3}
\left|
	\begin{array}{c c c }
	I'-g_{0}^{2} & -g_0 & -g_0\\	
	-g_0 &  \frac{\tilde z(t_0^*)^2}{1-\tilde z(t_0^*)^2} & \frac{\tilde z(t_0^*)\tilde z(t_0)}
  {1-\tilde z(t_0^*)\tilde z(t_0)}\\
	-g_0 &\frac{\tilde z(t_0^*)\tilde z(t_0)}{1-\tilde z(t_0^*)z(t_0)} & \frac{\tilde z(t_0)^2}{1-\tilde z(t_0)^2}  \\
	\end{array}\right|\leq 0.
\eeq
The determinant can be generalized in a straightforward way to include an additional interior point,  such as  the CT point for the scalar form factor. The corresponding domains are given in Figs. \ref{fig:dom1}-\ref{fig:dom3}. 
On the real axis, the figures indicate the points where double zeros are excluded. The small domains are obtained without 
including information on the phase and modulus on the unitarity cut. As discussed in Sec. \ref{sec:unib}, this case is obtained 
formally by replacing $\tin$ by $t_+$.
  
\vspace{0.4cm}

\vspace{0.5cm}

\begin{figure}[htb]
 	 \includegraphics[width = 7.1cm]{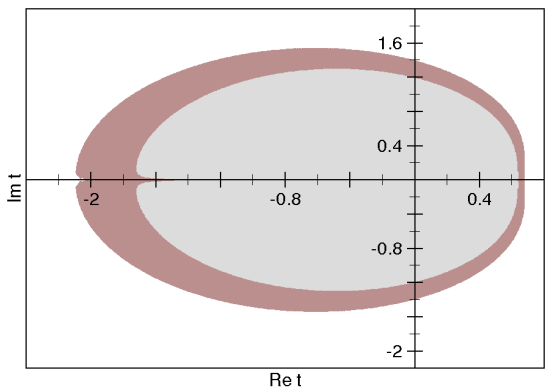}
	\caption{Domains where zeros of the vector form factor are excluded,  derived from (\ref{eq:detz3}), using $f_+(0)=0.962$ and phase and modulus information up to $\tin= 1\, {\rm GeV}^2$, when  $I'_+$ is varied within the errors.  The small (large) domain is 
obtained with the maximum (minimum) value of $I'_+$ given in (\ref{eq:Ip}).}
	\label{fig:domvecmaxmin}
\end{figure}

\begin{figure}[htb]
 	 \includegraphics[width = 7.1cm]{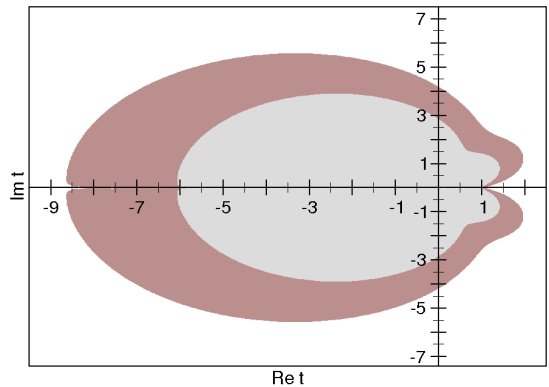}
	\caption{Domains where zeros of the scalar form factor are excluded, derived 
using $f_+(0)=0.962$, $f_0(\Delta_{K\pi})=1.193$, and phase and modulus information up to $\tin= 1\, {\rm GeV}^2$, when $I'_0$ is varied within the errors.  The small (large) domain is 
obtained with the maximum (minimum) value of $I'_0$ given in (\ref{eq:Ip}).}
	\label{fig:domscamaxmin}
\end{figure}

As in the case of the real zeros, the issue of the stability of the exclusion
regions of complex zeros is an important one against variations of the
inputs.  Here, too, the main uncertainty stems from those of $I'_+$ and
$I'_0$.  The results of these are shown in Figs. \ref{fig:domvecmaxmin}
and \ref{fig:domscamaxmin}.
\vspace{0.4cm}
\vspace{0.4cm}

We emphasize that our method is able to give in a rigorous way the regions 
where zeros are excluded, but says nothing about the remaining regions. 
Thus, we cannot answer the question whether the zeros are excluded 
everywhere, or a zero must exist. We recall that we  applied a formalism that
exploits a necessary condition, (\ref{eq:hI}), which follows from (\ref{eq:I}), the Fermi-Watson theorem, 
and the knowledge of the modulus in the elastic region. Therefore, the violation of this condition is sufficient to ensure
that the zero is not allowed.

\subsection{Phenomenological consequences}\label{sec:cons}
Our results show that 
the zeros are excluded in a rather large domain at low energies.
 This provides confidence in the semiphenomenological 
analyses based on Omn\`es representations, like those proposed in \cite{Bernard:2006gy,  Bernard:2009zm}, which assume 
that the zeros 
are absent.
 Indeed, a zero of the scalar form factor at $-0.1\,\mbox{GeV}^2$, which would distort the shape at low energies 
\cite{Bernard:2009zm}, is ruled out by
our results. On the other hand, as shown in \cite{Bernard:2009zm}, a real zero at $-1\,\mbox{GeV}^2$ leads already to a 
phenomenological form factor very similar to one with no
zeros. This value is close to the extremity of $-0.91\,\mbox{GeV}^2$ of the range without zeros for the scalar form factor, 
and the inclusion of the  additional constraint at the CT point rules out
a zero even at $-1\,\mbox{GeV}^2$.  In the case of complex zeros as well, the allowed zeros are rather remote to produce 
visible effects: for instance, a pair of complex
zeros located at $t_0= (0.1 \pm\,  2 \mbox{i})\, \mbox{GeV}^2$, considered in \cite{Bernard:2009zm}, is 
ruled out by our results.

In the case of the vector form factor, the analysis made in \cite{Bernard:2009zm} using data from $\tau$ decay concludes that 
complex zeros cannot be excluded, due to the lack of information on the phase of the form factor in the inelastic region. In 
contrast, as shown in Fig. \ref{fig:dom1}, the formalism presented here leads without any assumptions to a rather large domain  
where complex zeros are excluded. The stability analysis shows that, although for the maximum values of $I_+^{\prime}$ and 
$I_0^{\prime}$ from (\ref{eq:Ip}) the domains do shrink, the implications for the phenomenological analyses are unchanged.

\section{Conclusions }
\label{Conclusion}
In this work we have studied  the shape of the scalar and vector
form factors in the $K_{\ell 3}$ domain,  crucial
for the determination of the modulus of the CKM matrix element $|V_{us}|$. 
We applied a formalism proposed in \cite{Caprini2000}, 
which develops the standard method of unitarity bounds by including in an optimal way the phase and the modulus on 
the elastic part of the 
 unitarity cut. The formalism is also suitable for including information
from soft-meson theorems at points inside the analyticity domain.

The work reported here is a detailed application of the 
techniques explored in ~\cite{Abbas:2009dz}, also wiring in the
constraint to $t_{in}$ explored in ~\cite{Abbas:2010jc}, to the
phenomenology of the kaon semileptonic decay.  It uses the powerful
modified formalism for unitarity bounds and constraints from low-energy
theorems.  It goes beyond explorations of ~\cite{AbAn} which does
not use phase and modulus information, or ~\cite{BC} which does not
use modulus information for the scalar form factors.  Here, the vector
form factor is analyzed using the modified formalism which has never
been done before.  The modified formalism is employed to isolate those
regions of the real energy line and complex energy plane where zeros
are forbidden.  Thus, this work represents a powerful application of
the theory of unitarity bounds, which relies not so much on experimental
information, but on theoretical inputs from perturbative QCD.  It provides 
a powerful consistency check on determinations of shape
parameters from phenomenology and experimental analyses.
It may be noted that in the context of $B\to D^* l\nu$ the dispersive bounds 
obtained in ~\cite{CLN} were successfully used by experimental
groups studying the decay. 

The method exploits analyticity and unitarity, but differs  in several respects from the usual dispersion relations. In applying these relations, some assumptions about the form factor above the inelastic threshold are  necessary. Moreover, in the Omn\`es-type representations it is assumed that the form factors have no zeros in the complex plane. No such assumptions are necessary in our approach. Instead, one exploits a dispersion relation for a QCD correlator, which is calculated perturbatively in the Euclidian region and is related by unitarity to the modulus squared  of the form factors in the Minkowskian region. Positivity of the spectral functions then leads  to an integral relation of the form (\ref{eq:I}) for the modulus squared of the form factor, which is the basic relation of the approach. From such a relation one cannot make definite predictions for the  values of the form factors or their derivatives, but only derive bounds on these values. On the other hand, a remarkable feature is that an arbitrary number of values can be included simultaneously, corresponding mathematically to the so-called Meiman problem. Thus, the formalism is very useful for finding correlations between the  values of the form factors at different points and for testing the consistency of inputs known from different sources on different regions of the complex plane.  

In this work, we have focused  on the phenomenological consequences of the formalism.  We have considered  the standard parameterization of the  scalar and vector
form factors in the $K_{\ell 3}$ physical region, based on the Taylor expansion at $t=0$, and  derived constraints on the coefficients of the expansion. The results for the slope and curvature are given in simple analytic form in Eqs. (\ref{eq:normvect})-(\ref{eq:normscCT}) for arbitrary values of $f_+(0)$ and $f_0(\Delta_{K\pi})$.
The numerical coefficients in these inequalities depend on the  dispersion relation satisfied by the QCD correlators, their perturbative expressions, and the input phase and modulus on the elastic part of the cut. The sensitivity of the coefficients to the uncertainty of the input is quite low, as shown in \cite{Abbas:2009dz}. 

The constraints (\ref{eq:normvect})-(\ref{eq:normscCT}) can be used in the experimental fits with quadratic polynomials, 
or for testing {\em a posteriori} the consistency of the fitted parameters   with theoretical information 
 from regions outside the $K_{\ell 3}$ range. 
For illustration,  the small ellipses in Figs. \ref{fig:vect1}-\ref{fig:scal2} represent 
these domains for the default input (\ref{eq:f0+0}). The allowed values also satisfy 
the standard bounds and the phase condition in the elastic region, being included in the larger ellipses in the figures. 

 A more general condition, which correlates the coefficients of a cubic Taylor expansion,  
is given in Eqs. (\ref{eq:c3vect}) and (\ref{eq:c3sc}) for the vector and scalar form factors, respectively.  These relations can be used either to estimate  the  truncation error of the quadratic expansions, or for constraining the fits based on a cubic parameterization.

We have worked in the isospin limit, but have also given
a brief discussion of isospin breaking effects. The relation (\ref{eq:normscCTiso}), obtained using the mass of the neutral pion instead of the charged one, allows one to correlate the  isospin corrections in the slope and curvature of the scalar $K^+\pi^0$ form factor to those at $t=0$ and $t=\Delta_{K\pi}$.

We have also considered two other applications of the formalism: in (\ref{eq:ctctbar}) we  have derived  
a relation  between the values of the scalar form factor at the first and the second CT points, for an arbitrary $f_+(0)$. 
This represents a nontrivial analyticity constraint for the predictions beyond the SM suggested in \cite{Bernard:2006gy,Deschamps:2009rh,Passemar}. We have studied also the possible zeros  of the form factors, and have shown  that they are excluded in a rather large domain at low energies for both the vector and the scalar form factors. The results support the recent dispersive representations of the Omn\`es type, which assume that the zeros are absent. 

Finally, we  point out that alternative expansions, for instance, in powers of the variable  $z$ defined in Sec. \ref{sec:zparam}, may be more convenient from the point of view of convergence than the standard Taylor parameterizations (\ref{eq:taylor}). Constraints on the coefficients of such expansions can be derived using similar techniques, and will be investigated in a future work.

\vskip0.2cm

\noindent{\bf Acknowledgement:} 
BA thanks the Department of Science and Technology, Government of India, and the Homi Bhabha Fellowships Council for 
support. IC acknowledges support from  CNCSIS in the Program Idei, Contract No.
464/2009, and from the National Authority for Scientific Research , Project No. PN-09370102.  We thank S. Ramanan for
discussions and collaboration at an early stage of this project,
and V. Bernard and E. Passemar for discussions and for their
careful reading of and comments on the manuscript.


\end{document}